\documentclass[final,5p,times,twocolumn]{elsarticle}

 \usepackage{graphicx}

\usepackage{rotating,color,subfigure,amssymb}
\usepackage[T1]{fontenc}
\usepackage{alphalph}\renewcommand\theaffn{\alphalph{\arabic{affn}}}

\journal{Physics Letters B}

\begin{document}

\begin{frontmatter}

\title{Exclusive Measurement of the $\eta\to\pi^+\pi^-\gamma$ Decay }

\author[IKPUU]{The WASA-at-COSY Collaboration\\[2ex] P.~Adlarson}
\author[Erl]{C.~Adolph}
\author[ASWarsN]{W.~Augustyniak}
\author[IPJ,IKPJ,JCHP]{W.~Bardan}
\author[PITue]{M.~Bashkanov}
\author[IPJ]{T.~Bednarski}
\author[MS]{F.S.~Bergmann}
\author[ASWarsH]{M.~Ber{\l}owski}
\author[IITB]{H.~Bhatt}
\author[HISKP]{K.--T.~Brinkmann}
% ...

\address[IKPUU]{Division of Nuclear Physics, Department of Physics and 
 Astronomy, Uppsala University, Box 516, 75120 Uppsala, Sweden}
\address[Erl]{Physikalisches Institut, Friedrich--Alexander--Universit\"at 
 Erlangen--N\"urnberg, Erwin--Rommel-Str.~1, 91058 Erlangen, Germany}
\address[ASWarsN]{Department of Nuclear Reactions, The Andrzej Soltan 
 Institute for Nuclear Studies, ul.\ Hoza~69, 00-681, Warsaw, Poland}
\address[IPJ]{Institute of Physics, Jagiellonian University, ul.\ Reymonta~4, 
 30-059 Krak\'{o}w, Poland}
\address[IKPJ]{Institut f\"ur Kernphysik, Forschungszentrum J\"ulich, 52425 
 J\"ulich, Germany}
\address[JCHP]{J\"ulich Center for Hadron Physics, Forschungszentrum J\"ulich, 
 52425 J\"ulich, Germany}
\address[PITue]{Physikalisches Institut, Eberhard--Karls--Universit\"at 
 T\"ubingen, Auf der Morgenstelle~14, 72076 T\"ubingen, Germany}
\address[MS]{Institut f\"ur Kernphysik, Westf\"alische Wilhelms--Universit\"at
 M\"unster, Wilhelm--Klemm--Str.~9, 48149 M\"unster, Germany}
\address[ASWarsH]{High Energy Physics Department, The Andrzej Soltan Institute
 for Nuclear Studies, ul.\ Hoza~69, 00-681, Warsaw, Poland}
\address[IITB]{Department of Physics, Indian Institute of Technology Bombay, 
 Powai, Mumbai--400076, Maharashtra, India}
\address[HISKP]{Helmholtz--Institut f\"ur Strahlen-- und Kernphysik, 
 Rheinische Friedrich--Wilhelms--Universit\"at Bonn, Nu{\ss}allee~14--16, 
 53115 Bonn, Germany}
% ...

\fntext[fnec]{present address: INFN, Laboratori Nazionali di Frascati, Via E. 
 Fermi~40, 00044 Frascati (Roma), Italy}
\fntext[fncp]{present address: Fachbereich Physik, Bergische Universit\"at 
 Wuppertal, Gau{\ss}str.~20, 42119 Wuppertal, Germany}
\fntext[fntt]{present address: Albert Einstein Center for Fundamental Physics,
 Fachbereich Physik und Astronomie, Universit\"at Bern, Sidlerstr.~5, 
 3012 Bern, Switzerland}

\cortext[coau]{Corresponding author }

\begin{abstract}

An exclusive measurement of the decay $\eta\to\pi^+\pi^-\gamma$ has been performed at the WASA facility at
COSY. The $\eta$ mesons were produced in the fusion reaction pd$\to ^3$He X at a proton beam momentum of
1.7~GeV/c. Efficiency corrected differential distributions have been extracted based on 13960$\pm$140 events
after background subtraction. The measured pion angular distribution is consistent with a relative {\it
p}-wave of the two-pion system, whereas the measured photon energy spectrum was found at variance with the
simplest gauge invariant matrix element of $\eta\to\pi^+\pi^-\gamma$. A parameterization of the data can be
achieved by the additional inclusion of the empirical pion vector form factor multiplied by a first-order
polynomial in the squared invariant mass of the $\pi^+\pi^-$ system.
\end{abstract}


\end{frontmatter}

\section{Introduction}\label{sec:intro}

The $\eta$ meson plays a special role in understanding low--energy Quantum Chromo Dynamics (QCD). Chiral
symmetry, its realization in hadron physics at low energies and the role of explicit chiral symmetry breaking
due to the masses of the light quarks $(u,d,s)$ can be investigated using $\eta$ decays. This work focuses on
the anomalous sector of QCD, which is manifested in the radiative decays of the $\eta$ meson. 

The radiative decay $\eta\to\pi^+\pi^-\gamma$ is the fourth strongest decay mode of the $\eta$ meson with a
branching ratio of 4.60~$\pm$~0.16\%~\cite{Nakamura:2010zzi}. Conservation of charge conjugation and angular
momentum including parity constrain the dynamics of the decay products. The photon and the $\eta$ meson are
eigenstates of the charge symmetry transformation with the eigenvalues C = --1 and C = +1, respectively.
Therefore, due to C invariance the $\pi^+\pi^-$ system must have C = --1. To ensure C invariance, the orbital
angular momentum $L$ between the two pions must be odd. All involved particles have negative parity.
Consequently, parity invariance demands that the orbital angular momentum $L'$ between the photon and the
two-pion system must also be odd. Finally, total angular momentum conservation incorporating the fact that
the intrinsic spin of the photon is unity leads to the requirement $L=L'$. Thus, the lowest partial waves
which contribute are {\it p}-waves. Presumably, higher partial waves with $L \ge 3$ are not very important.

Radiative decays of the $\eta$ meson are driven by the chiral anomaly of QCD. The effects of the anomaly have
been summarized by Wess and Zumino in an effective Lagrangian~\cite{Wess:1971}. As shown by Witten, this
Lagrangian is an essential part of effective field theories, because it is necessary in order to correctly
incorporate the parity transformation of QCD~\cite{Witten:1983}. At the chiral limit of zero momentum and
massless quarks the decay $\eta\to\pi^+\pi^-\gamma$ is determined by the box anomaly term of the
Wess-Zumino-Witten Lagrangian, which describes the direct coupling of three pseudoscalar mesons and a photon.
The dynamic range of the decay is limited by two pion rest masses and the $\eta$ mass, $4m_\pi^2 \leq
s_{\pi\pi} \leq m_\eta^2$, and is, thus, far from the chiral limit. As a consequence, the decay rate
calculated from the box anomaly term at the tree level is smaller by a factor of two compared to the measured
value. Higher order terms of the chiral Lagrangian have to be taken into account to achieve a correct
description of the decay $\eta\to\pi^+\pi^-\gamma$. Calculations at the one-loop level show an improved
agreement between experiment and theory~\cite{Bijnens:1989}. But there remains, however, a significant
difference. Several efforts have been made to include final state interactions by unitarized extensions to the
box-anomaly term, e.g. a momentum dependent Vector Meson Dominance (VMD) model~\cite{Picciotto:1991}, the
Hidden Local Symmetry model~\cite{Benayoun:2003}, an Omnes-function which accounts for the {\it p}-wave pion
scattering phase shift~\cite{Venugopal:1998fq,Holstein:2001}, and a Chiral Unitary
approach~\cite{Borasoy:2004}.

To test the validity of the different models, not only the decay rate but also differential distributions of
the Dalitz plot variables need to be compared with experimental data. For this purpose it is useful to
parameterize the Dalitz plot in terms of the photon energy $E_\gamma$ in the rest frame of the $\eta$ meson
and the angle~$\theta$ of the $\pi^+$ relative to the photon in the pion-pion rest frame. E$_\gamma$ is
related to the squared invariant mass of the pion pair $s_{\pi\pi}$ according to
\begin{equation}
  E_{\gamma} = \frac{1}{2}\left( m_{\eta} - \frac{s_{\pi\pi}}{m_{\eta}} \right) \ .
\end{equation}

The photon energy distribution has been subject of only a few measurements forty years
ago~\cite{Crawford:1966,Cnops:1968,Gormley:1970,Layter:1973}. The results of the two statistically most
significant publications~\cite{Gormley:1970,Layter:1973} are presented without acceptance corrections.  
Instead, the models which have been used for the interpretation of the data were folded with the acceptance
and the results seem to be inconsistent~\cite{Benayoun:2003, Borasoy:2007}.

Due to the limitations of the currently available experimental data the potential of the decay
$\eta\to\pi^+\pi^-\gamma$ to provide insight to the anomalous sector of QCD cannot be fully exploited. In
order to perform compelling tests of the box anomaly and its higher order terms in Chiral Perturbation Theory
as well as of the non-perturbative extensions of the box-anomaly terms a new measurement with higher precision
is called for.

\section{The Experiment}
\label{section:exp_setup}

The results presented in this paper are based on a measurement with the WASA detection system~\cite{Adam:2004}
installed at the Cooler Synchrotron COSY~\cite{Maier:1997zj} at the Forschungszentrum J\"ulich in Germany. A
pellet target system produces small spheres of frozen hydrogen or deuterium which interact with the ion beam
of the accelerator. The interaction region is surrounded by a central detector covering scattering angles from
20 to 169 degrees. It consists of a straw tube drift chamber, which is operated in the magnetic field of
0.85~T provided by a superconducting solenoid for the momentum reconstruction of charged particles, an
electromagnetic calorimeter to measure energies of charged as well as neutral particles and thin plastic
scintillators to discriminate charged and neutral particles already at the trigger level. Energy loss patterns
in the calorimeter and the plastic scintillators allow to identify charged particles. For the identification
and reconstruction of particles emitted at polar angles from 3 to 18 degrees a forward detector is used. While
track coordinates are measured precisely by a straw tube drift chamber, the kinetic energies of the ejectiles
are reconstructed from the signals in plastic scintillators of different thickness, using the $\Delta$E-E
method.

The $\eta$ mesons have been produced in the fusion reaction pd$\to ^3$He X at a proton beam momentum of
1.7~GeV/c. This corresponds to an excess energy of 60~MeV in the center of mass for the reaction pd$\to ^3$He
$\eta$ at a cross section of 0.412~$\mu$b~\cite{Bilger:2002}. At the trigger level events with one track in
the forward detector with a high energy deposit in the thin scintillator detectors close to the exit window of
the scattering chamber have been accepted. This condition selects $^3$He ions without bias on the $\eta$ decay
system.

\begin{figure}[htb]
  \centerline{\includegraphics[width=0.5\linewidth]{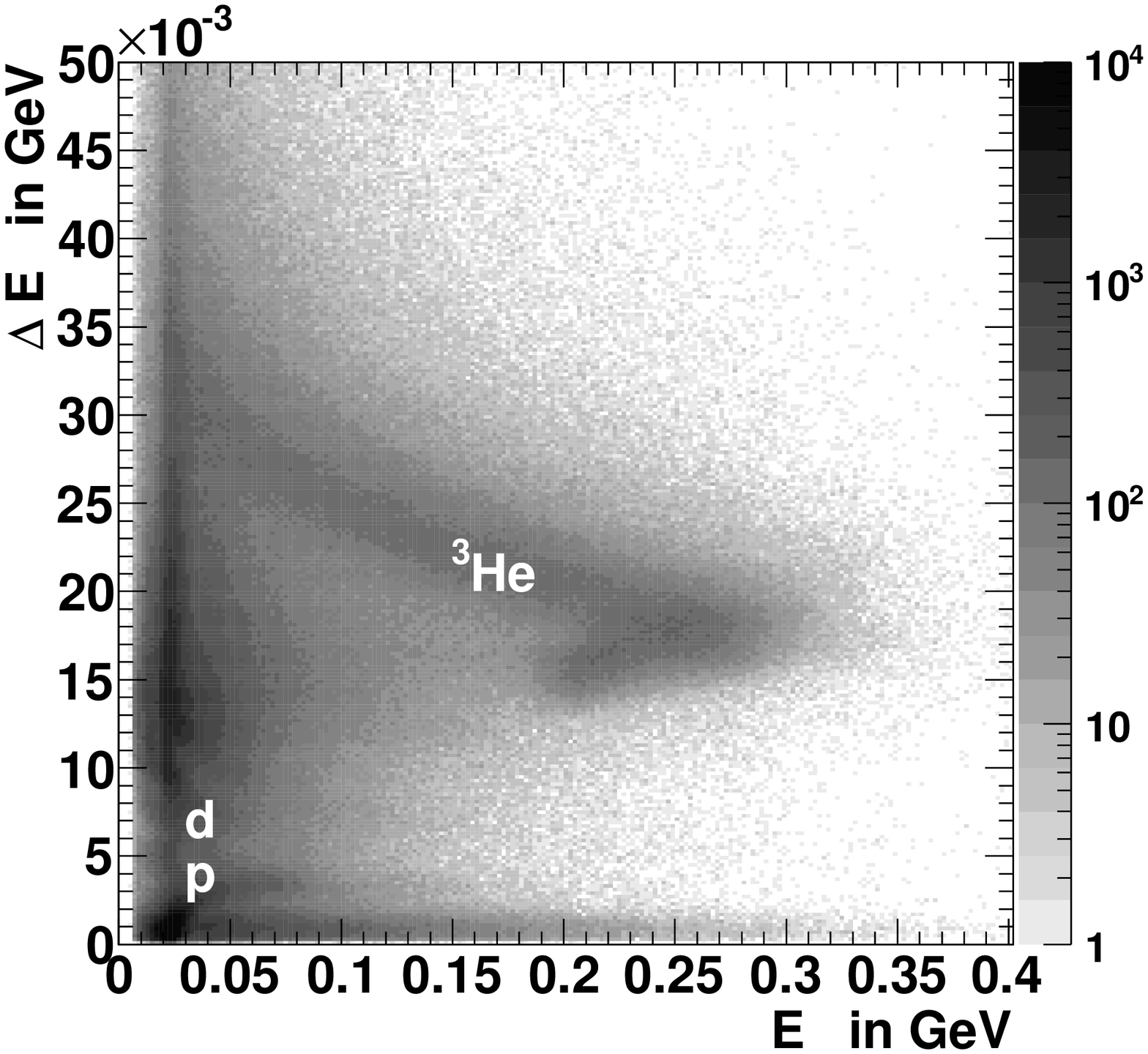}%
              \includegraphics[width=0.5\linewidth]{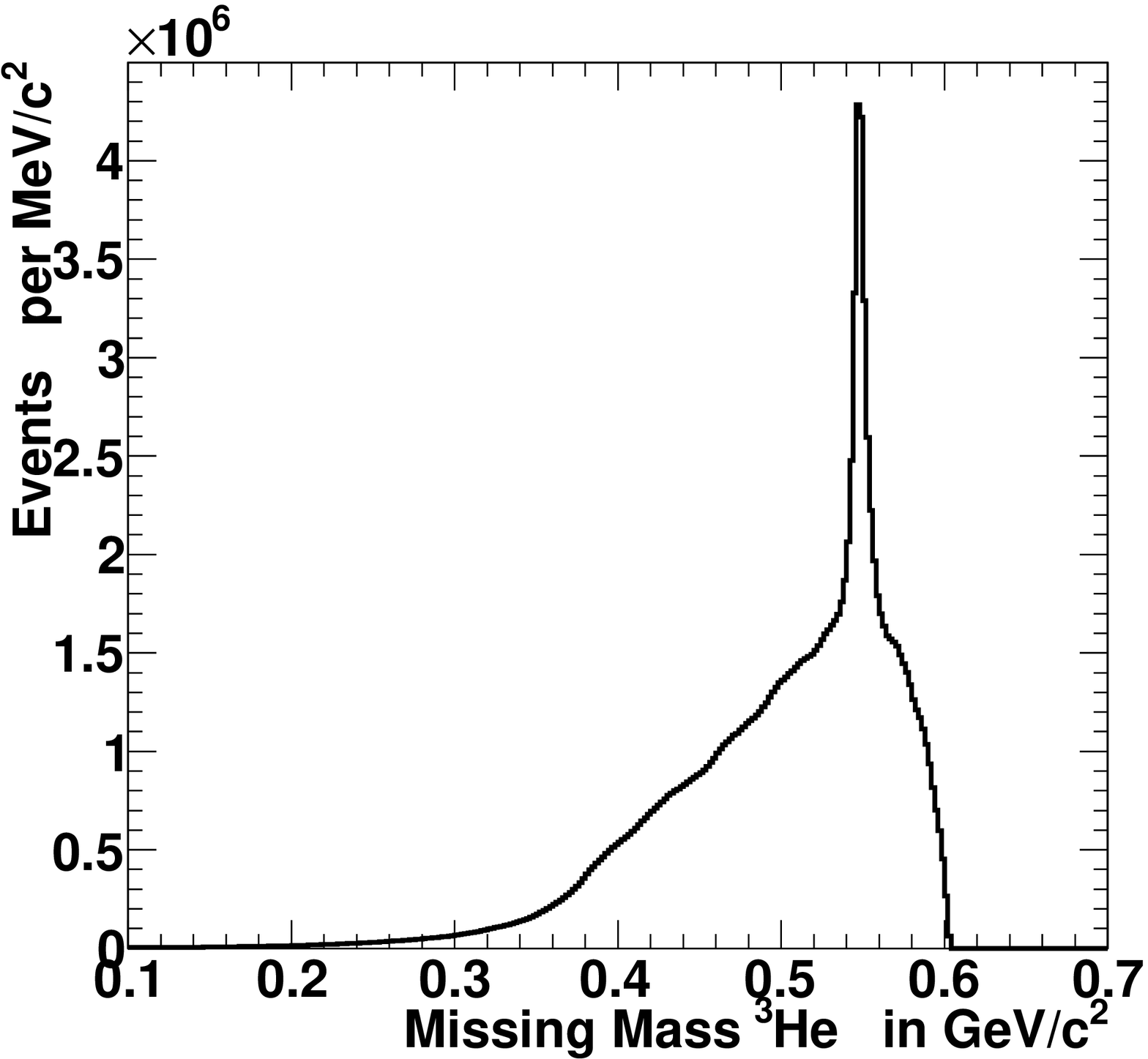}}
  \caption{\label{fig:HeMM}(Left) $\Delta$E-E plot for tracks emitted at scattering angles between 3 and 18
degree. $^3$He is well separated from protons and deuterons. (Right) Missing mass of $^3$He for all events in
the $^3$He band. The number of events in the peak above the background is 1.1$\cdot 10^{7}$.}
\end{figure}

Fig.~\ref{fig:HeMM} shows the energy loss correlations of all tracks in the forward detector. The structure
attributed to $^3$He is clearly visible and can be easily selected. It is well separated from the structures
of protons and deuterons, which are strongly suppressed due to the trigger conditions. The right panel of
Fig.~\ref{fig:HeMM} shows the inclusive missing mass distribution of $^3$He. A pronounced peak at the $\eta$
meson mass is visible. It contains 1.1$\cdot10^7$ tagged mesons on top of a continuous background which
originates from direct multi-pion production. The cross sections of the relevant background processes
\mbox{pd$\to ^3$He~$\pi^+\pi^-$} and \mbox{pd$\to ^3$He~$\pi^+\pi^-\pi^0$} are unknown at the center-of-mass
energy of this measurement. However, they can be estimated, by extrapolation of results at lower
energies~\cite{Bashkanov:2005fh}, to be about 5~$\mu$b and 0.4~$\mu$b, respectively. The missing mass spectrum
in the right panel of Fig.~\ref{fig:HeMM} illustrates that the direct reactions contribute to the background
for $\eta$ decays only in a range of a few MeV/c$^2$ around the $\eta$ meson mass. This remaining background
is subtracted in a model independent way, bin by bin from the final spectra.

\section{Data Analysis}

For the selection of the final state, one charged particle, which is identified as $^3$He, is required in the
forward detector. In the central detector, two charged particle tracks of opposite curvature and one neutral
particle track with an energy deposit in the calorimeter of at least 20~MeV are required in addition. Both
charged particles are identified as pions. Background events which are picked up by the selection rules stem
mainly from charged multi-pion production and other $\eta$ decay modes.

\begin{figure}[htb]
  \includegraphics[width=0.49\linewidth]{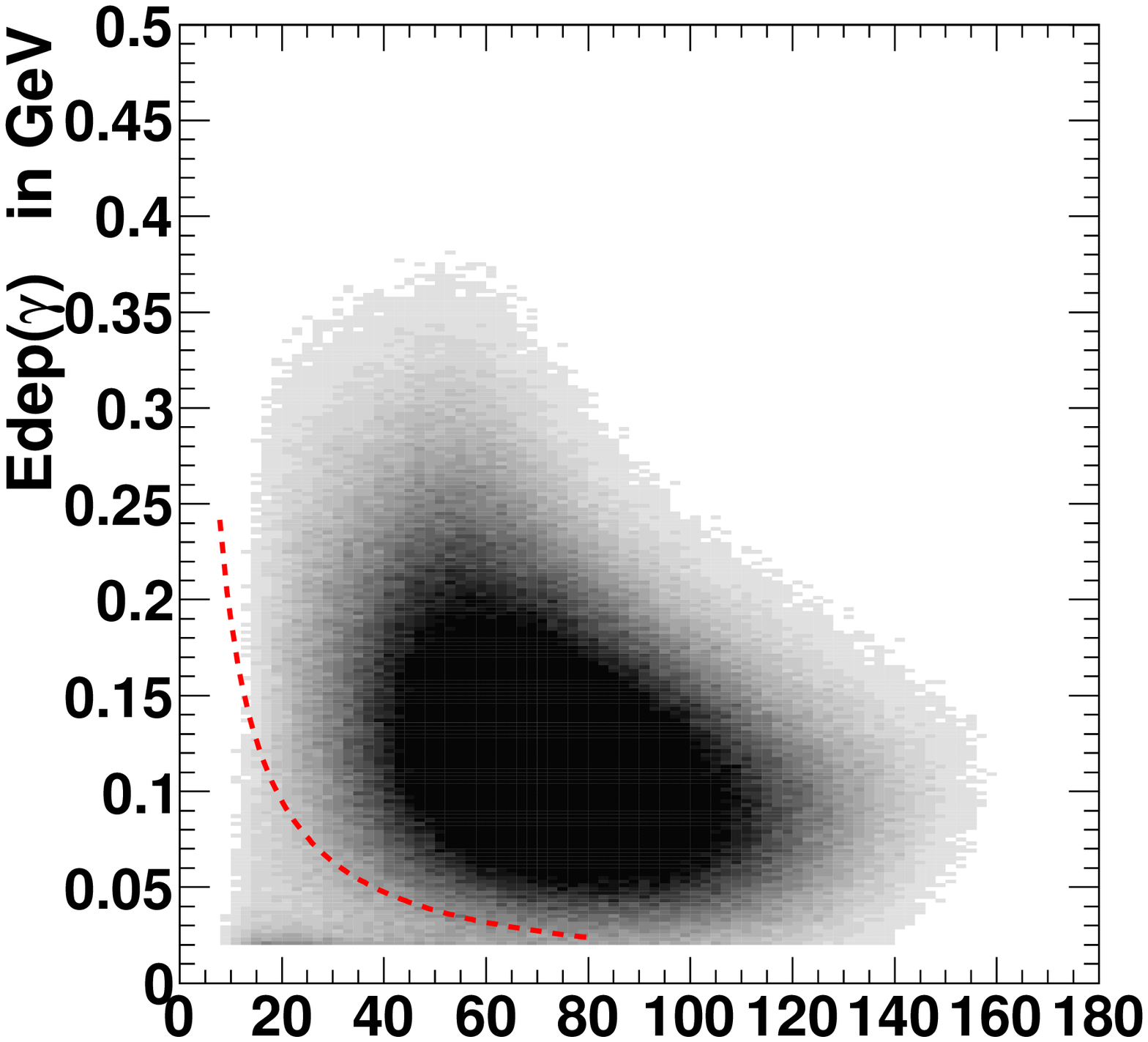}%
  \includegraphics[width=0.49\linewidth]{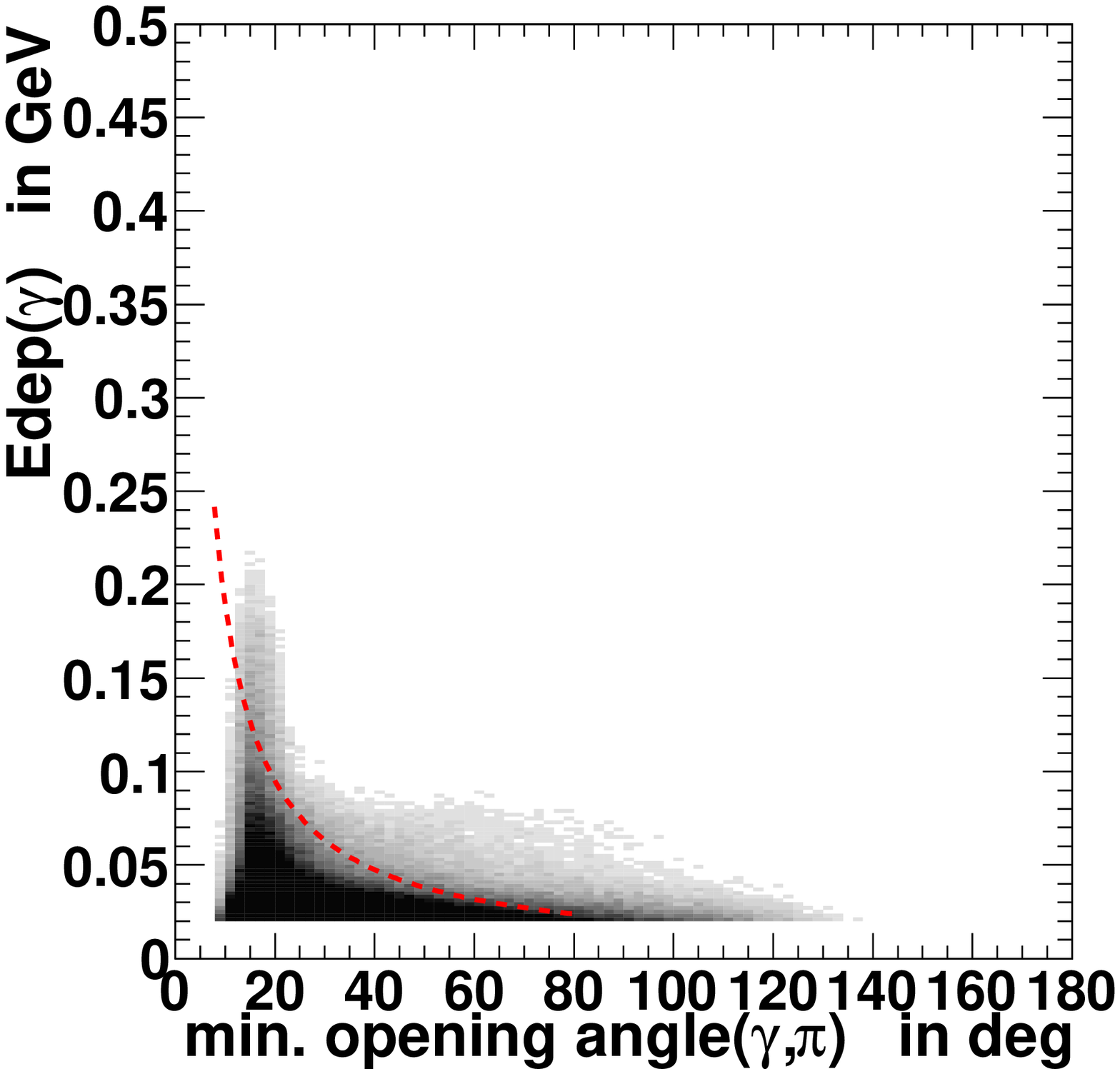}
  \caption{\label{fig:split} Correlation plot for laboratory photon energy vs charged track photon opening
angle (Monte Carlo) (left:) signal, (right:)  $pd\to^3$He$\pi^+\pi^-$ reaction. The graphical cut to suppress 
\mbox{pd$\to  ^3$He~$\pi^+\pi^-$} reaction is shown.}
\end{figure}
Two-pion production is expected to form the largest background contribution in the selected data due to the
large cross section and the possibility of cluster splitoffs in the calorimeter, which would fake the signal
of a photon candidate in the actually photon free final state. Splitoffs are characterized by a small  energy
deposit and a small distance to either of the pion candidates in the calorimeter. Therefore, this background
contributes predominantly to the low end of the reconstructed $E_\gamma$ spectrum. In order to increase the
signal to background ratio in this region a correlated condition is imposed on the energy deposits of the
photon candidates and the distance between the shower positions of photon and pion candidates in the
calorimeter. The latter is measured by the opening angle between the reconstructed cluster positions. In
Fig.~\ref{fig:split} the condition is shown as graphical cut with a dashed curve. A strong enhancement is
observed at low photon energies and small opening angles between photon and pion candidates from the
contribution of the \mbox{pd$\to ^3$He~$\pi^+\pi^-$} reaction. The cut used is the best compromise between a
high signal-to-background ratio and a large reconstruction efficiency of the signal channel. The contribution
from two-pion production is reduced by more than 55\% by rejecting all events below the curve.

\begin{figure}[htb]
  \centerline{\includegraphics[width=\linewidth]{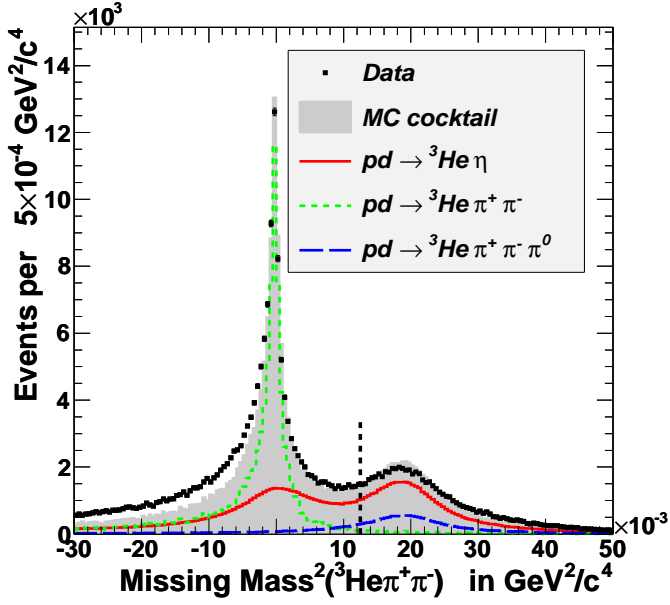}}	
  \caption{\label{fig:back}The $\pi^0$ signal in the squared missing mass of the $^3$He~$\pi^+\pi^-$ system. 
The dashed vertical line corresponds to a squared missing mass of 0.0125~GeV$^2$/c$^4$.}
\end{figure}

Final states with three pions contribute to the background only if one photon of the $\pi^0$ decay remains
undetected. The $\pi^0$ can be identified from the distribution of the squared missing mass of the
\mbox{$^3$He$\,\pi^+\pi^-$} system, as demonstrated in Fig.~\ref{fig:back}. Here, the experimental missing
mass spectrum is reproduced with Monte Carlo distributions of the the signal ($\eta\to\pi^+\pi^-\gamma$) and
background ($\eta\to\pi^+\pi^-\pi^0, \eta\to e^+ e^-\gamma$) contributions from $\eta$ decays as well as
direct two- and three-pion production. The cross sections of the  direct pion processes were fitted using
simultaneously the \mbox{$^3$He$\,\pi^+\pi^-$} missing mass distribution and the \mbox{$^3$He} missing mass
for the selected data sample. The result is in agreement with the expectations~\cite{Bashkanov:2005fh}. The
contributions of the different $\eta$ meson decay channels are fixed by the known branching
ratios~\cite{Nakamura:2010zzi}. The discrepancy between Monte Carlo and experiment for negative masses might
be attributed to the unknown production mechanisms of the direct processes. In the simulations isotropic phase
space population has been assumed. The discrepancy disappears when the contribution of the direct production
is subtracted bin-by-bin, as discussed later. By rejecting events with a squared missing mass value larger
than 0.0125~GeV$^2$/c$^4$, as indicated with a vertical dashed line in Fig.~\ref{fig:back}, $73\%$ of the
background from the \mbox{$^3$He$\,\pi^+\pi^-\pi^0$} final states in the mass region of the $\eta$ meson is
removed. 

\begin{figure}[htb]
    \centerline{\includegraphics[width=\linewidth]{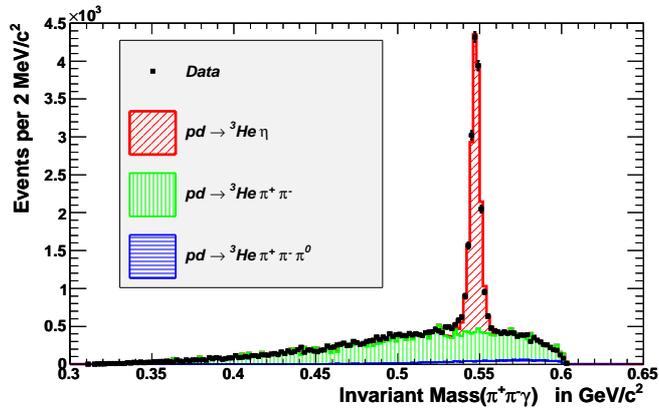}}
    \caption{\label{fig:kfit}The invariant mass spectrum of the fitted $\pi^+\pi^-\gamma$ system with a
probability above 10\%. The Monte Carlo cocktail consists of the signal ($\eta\to\pi^+\pi^-\gamma$) and
background contributions ($\eta\to\pi^+\pi^-\pi^0, \eta\to e^+ e^-\gamma$) from $\eta$ decays (diagonally
hatched) as well as direct two-pion (vertically hatched) and three-pion (horizontally hatched) production.}
\end{figure}

Additional suppression of background is achieved by a kinematic fit of the complete final state to the
reaction hypothesis \mbox{pd$\to^3$He~$\pi^+\pi^-\gamma$}, using four-momentum conservation as the only fit
constraint. The uncertainties of the kinematic variables at the input to the fit have been extracted
depending on energy and angle from a GEANT Monte Carlo simulation, tuned to match the experimental
resolutions. After the fit, all events with a probability of less than 10\% are rejected. The invariant mass
distribution of the fitted $\pi^+\pi^-\gamma$ system is shown for the remaining events in Fig.~\ref{fig:kfit}.
The condition on the probability distribution additionally suppresses background. In particular a comparison
with Monte Carlo shows a reduction of the $^3$He$\,\pi^+\pi^-\pi^0$ background to the level of $10\%$ in the
selected events.

\begin{figure}[t]
  \centerline{\includegraphics[width=0.5\linewidth]{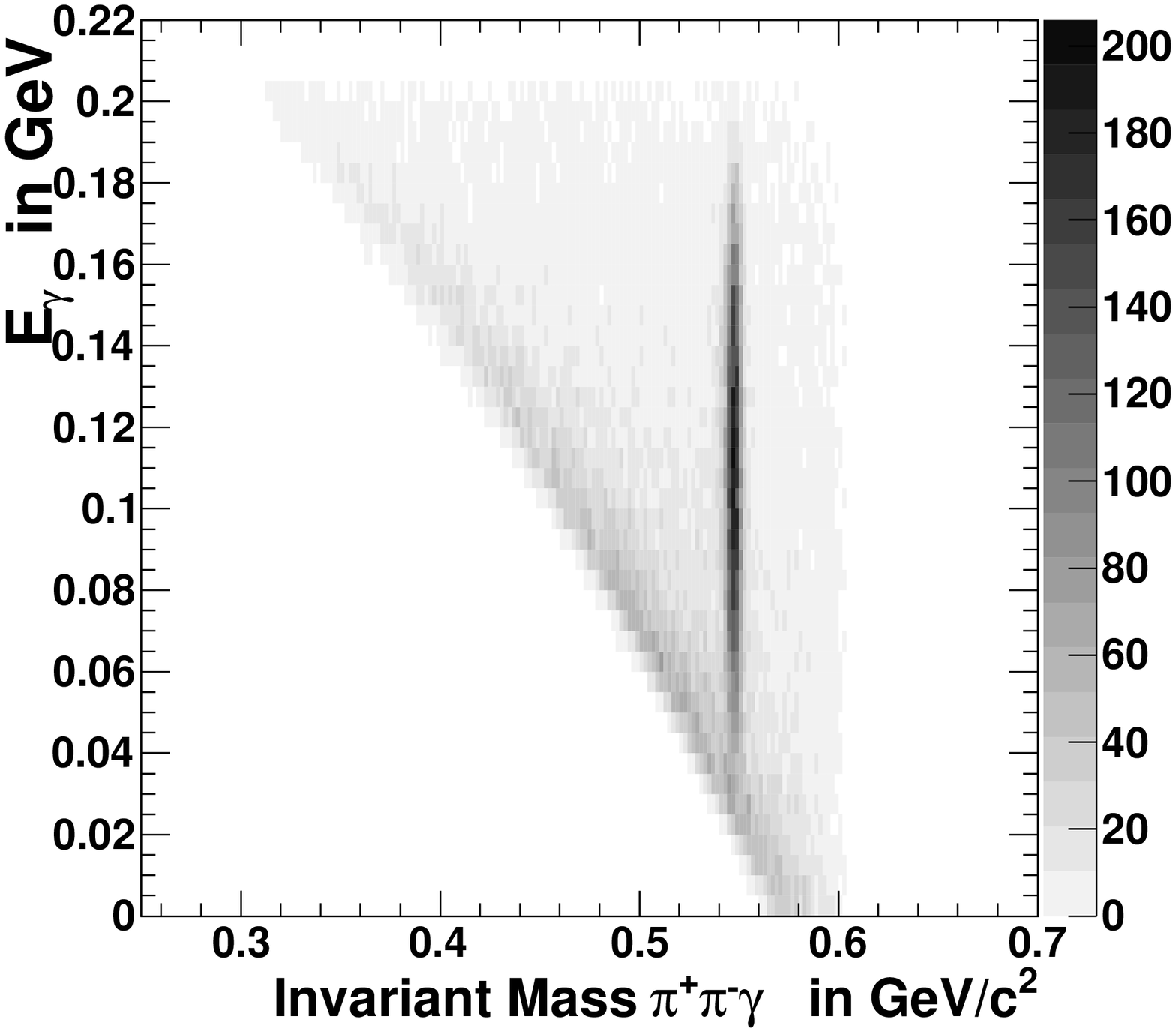}%
              \includegraphics[width=0.5\linewidth]{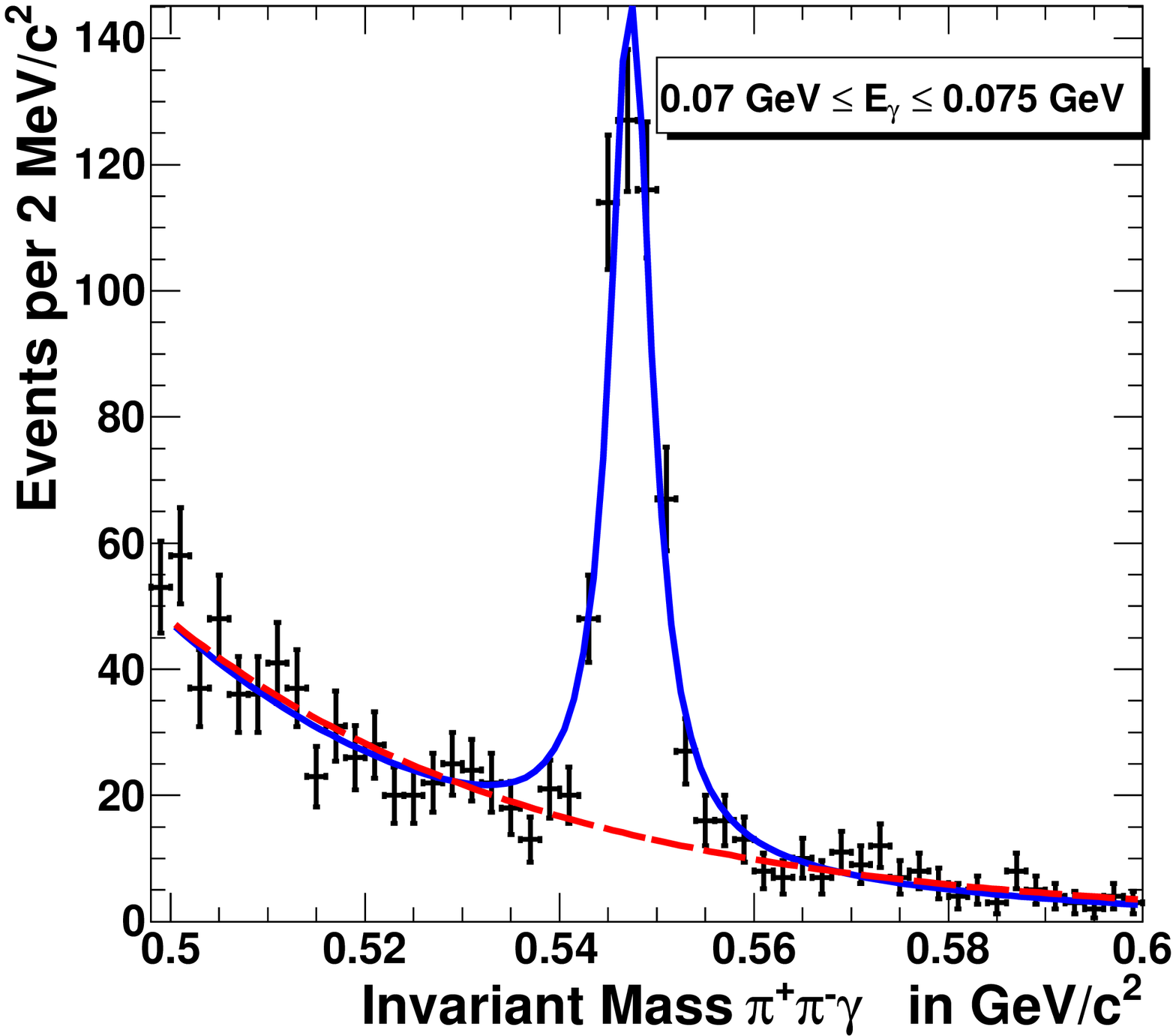}}
  \caption{\label{fig:bgsub} Background subtraction from the E$_\gamma$ distribution. \textbf{Left:}
Correlation of E$_\gamma$ and the invariant mass of the $\pi^+\pi^-\gamma$ system. \textbf{Right:}
Illustration of the background subtraction methods with the invariant mass spectrum corresponding to the
photon energy region \mbox{$70\leq E_\gamma \textrm{[MeV]}\leq 75$} : Determination of the background shape by
fitting signal and background (solid curve) and by excluding the signal range from the fit of the background
(dashed curve).}    
\end{figure}

The remaining background is subtracted bin by bin from the $E_\gamma$ and $\cos\theta$ distributions. The bin
size of 5~MeV for the photon energy and 0.1 for the pion angular distributions is chosen to reflect the
resolution achieved in the respective observables. The bin width of the angular distribution is similar to 
previous measurements~\cite{Gormley:1970, Layter:1973} the bin width of the $E_\gamma$ spectrum is smaller by
about a factor of two. The left panel in Fig.~\ref{fig:bgsub} shows the correlation between the photon energy
and the invariant mass  of the $\pi^+\pi^-\gamma$ system. Two structures can be seen in the plot. One at the
mass of the $\eta$~meson containing the signal events and another structure, showing a correlation of photon
energy and invariant mass, caused mainly by background from two-pion production. The invariant mass of the
$\pi^+\pi^-\gamma$ system is calculated for the events in each bin of the $E_\gamma$ and $\cos\theta$
distributions. This is illustrated for one bin of the $E_\gamma$ distribution in the right panel of
Fig.~\ref{fig:bgsub}. Since both, cross sections and differential distributions are unknown for multi-pion
production at the energy of this measurement, the individual mass spectra are fitted in two ways in order to
determine the amount of continuous background in the $\eta$  mass region. In the first approach a function
that is the sum of a Lorentzian for the signal and an exponential for the  background is used. In the second
method the background is fitted with an exponential function without making an assumption on the signal shape
by excluding the range of the signal peak from the fit. The excluded region was determined as the
$3\sigma$~region of a Gaussian fit to the peak of the $\eta$ meson. The individual results of the fits using
both methods are shown in the right panel of Fig.~\ref{fig:bgsub} with solid and dashed curves, respectively.
The average number of the background events from both fits is used to calculate the number of events from the
$\eta$~meson decay in the corresponding bin of the $E_\gamma$ or $\cos\theta$ distribution. 

The remaining background from the \mbox{$\eta\to\pi^+\pi^-\pi^0$} decay in the signal region is subtracted
using scaled Monte Carlo distributions. The scaling factors are determined by a fit of the experimental
spectrum of the squared missing mass of the \mbox{$^3$He~$\pi^+\pi^-$} system with Monte Carlo distributions
of the relevant$\eta$ decay modes, $\eta\to\pi^+\pi^-\gamma$ (signal), $\eta\to\pi^+\pi^-\pi^0$ and $\eta\to
e^+ e^-\gamma$ (background), after subtraction of the continuous background from multi-pion production. The
contribution of \mbox{$\eta\to\pi^+\pi^-\pi^0$} determined to be 7\%. This background is isotropic in
$\cos\theta$ and contributes to the E$_\gamma$ distribution in the energy region above 50~MeV with a maximum
at 120~MeV. The total statistics in the final distributions is $13960\pm140$ events of the
\mbox{$\eta\to\pi^+\pi^-\gamma$} decay. This is the largest number of events from an exclusive measurement of
this decay mode.

For acceptance corrections the general form of the squared matrix element for the $\eta\to\pi^+\pi^-\gamma$
decay
\begin{equation}\label{eq:sgime}
       |{\mathcal M}\,|^2 \sim |F(s_{\pi\pi})|^2\textrm{E}_\gamma^2 q^2 \sin^2(\theta)\qquad,
\end{equation}
with $q$ being the pion momentum in the pion-pion rest frame, and the form factor $F(s_{\pi\pi})$ according to
the VMD calculations in Ref.~\cite{Picciotto:1991} has been used. Using instead Monte Carlo distributions
based on the simplest gauge invariant matrix element ($F(s_{\pi\pi})=const.$) does not alter the experimental
result significantly. Thus, it can be concluded that systematic effects due to the applied form factors are
negligible. The acceptance varies smoothly as a function of $E_\gamma$ and $\cos\theta$. For photon energies
less than 10~MeV the acceptance becomes vanishingly small. In case of the angular distribution a reduced
acceptance is observed for small opening angles between each of the pions and the photon. The reduction of the
acceptance in both variables is found to be correlated. It is caused by the method of two-pion suppression
presented in Fig.~\ref{fig:back}, where a condition on the correlation of photon energy and opening angle
between pion and photon candidates is used.

\section{Results}
\begin{figure}[hbt]
    \centerline{\includegraphics[width=\linewidth]{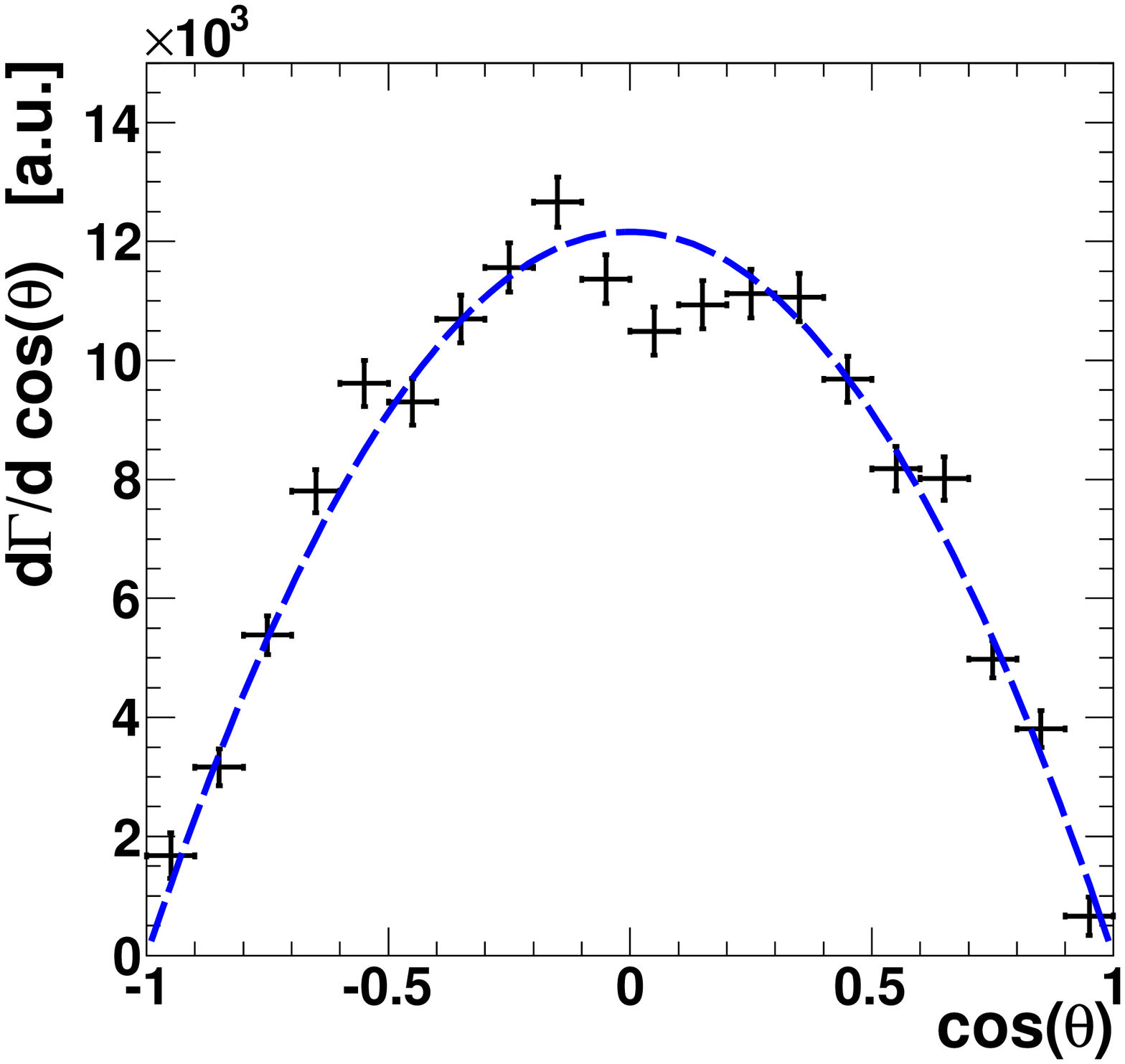}}
    \centerline{\includegraphics[width=\linewidth]{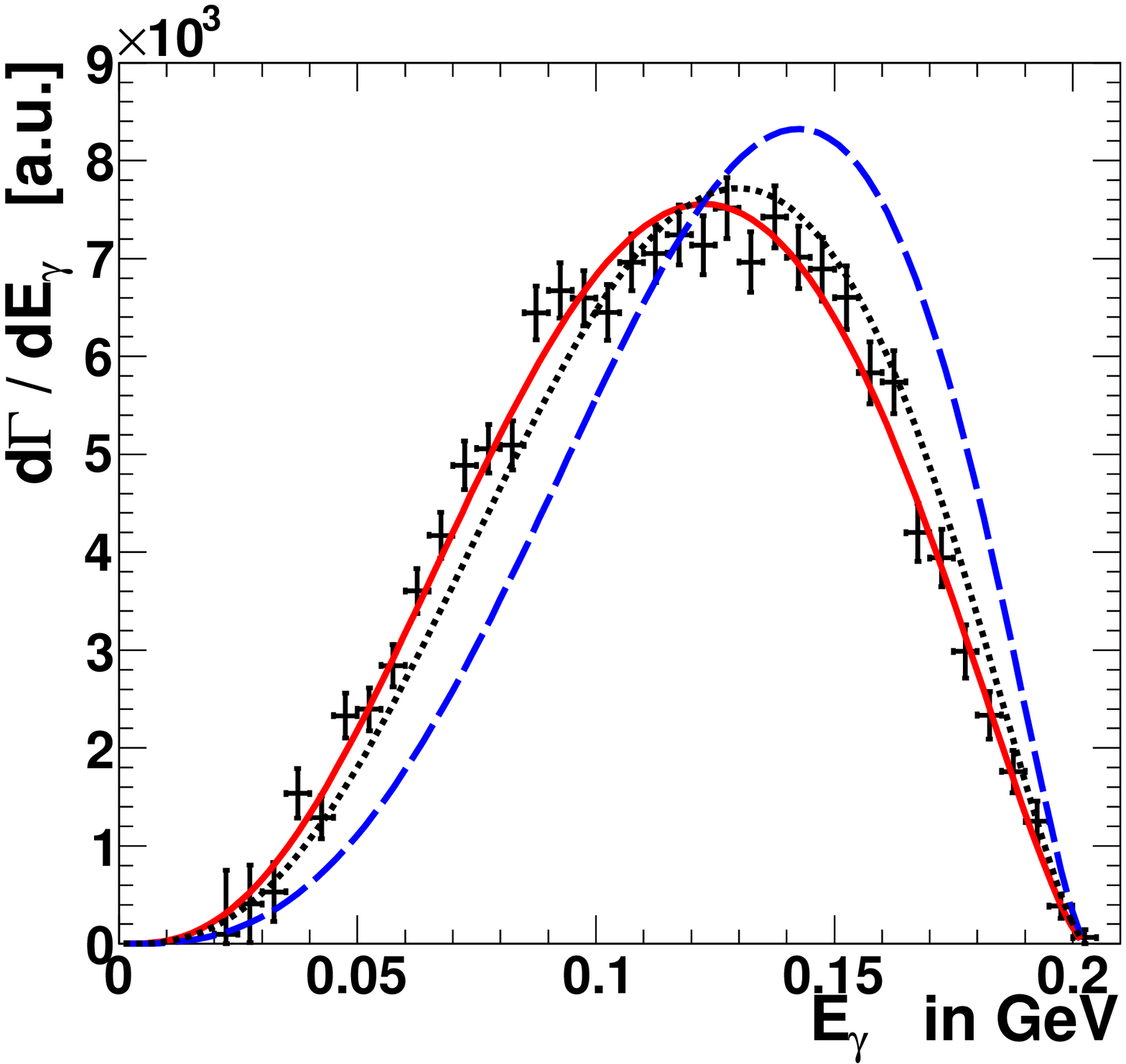}}
    \caption{\label{fig:final}The background subtracted and acceptance corrected angular distribution of the
pions (top) and the photon energy distribution (bottom), with error bars indicating the statistical
uncertainties. The angular distribution is compared with a relative {\it p}-wave of the pions (dashed curve).
The shape of the photon energy distribution is confronted with predictions of the square of the simplest gauge
invariant matrix element, Eq.~\ref{eq:sgime} (dashed curve), multiplied by the squared modulus of the pion
vector form factor $|F_V(s_{\pi\pi})|^2$ (dotted curve) and further multiplied by $(1+\alpha s_{\pi\pi})^2$,
the square of a real polynomial of first order, with its coefficient fitted to the data (solid curve). All
curves are normalized to the same integral.}
\end{figure}

Fig.~\ref{fig:final} shows the background subtracted and acceptance corrected photon energy and pion angular
distributions with the statistical errors. The E$_\gamma$ distribution is also given numerically in
Tab.~\ref{tab:EGfinal}.

In the upper panel of Fig.~\ref{fig:final} the final distribution of $\cos\theta$ is shown. It can be
described by \mbox{$\textrm{d}\,\sigma/\textrm{d}\cos\theta=A\cdot\sin^2(\theta)$}, as indicated by the dashed
curve. Thus, the measurement is consistent with the relative {\it p}-wave assumed in Eq.~\ref{eq:sgime}.

The photon energy distribution in the $\eta$ rest frame is shown in the lower panel of Fig.~\ref{fig:final}.
The line shape obtained from Eq.~\ref{eq:sgime}, normalized to the integrated rate, is given by the dashed
curve. It does not describe the experimental data, which in comparison are shifted significantly towards
lower energies. The observed disagreement confirms the findings of the previous
measurements~\cite{Crawford:1966, Cnops:1968, Gormley:1970, Layter:1973}.

In order to achieve a correct description of the photon energy spectrum in addition to the already properly
described angular distribution, Eq.~\ref{eq:sgime} can be multiplied by an energy dependent form factor
$\left|FF(s_{\pi\pi})\right|^2$. The origin of the deviation in the E$_\gamma$ distribution is predominantly
given by the $\pi\pi$ final state interaction in the vector channel. Unitarity and analyticity dictate that
this effect should be given by the pion vector form factor $F_V(s_{\pi\pi})$ multiplied by a polynomial
$P(s_{\pi\pi})$ that parameterizes contributions that do not contain the $\pi\pi$ unitarity cut. For a
detailed discussion about the multiplier $\left|FF(s_{\pi\pi})\right|^2 =
\left|F_V(s_{\pi\pi})P(s_{\pi\pi})\right|^2$ of Eq.~\ref{eq:sgime} see {\it
e.g.}~Refs.~\cite{Venugopal:1998fq,Holstein:2001,Truong:2001en,Stollenwerk:IP}. The pion vector form factor is
experimentally directly accessible via $e^+e^-\to \pi^+\pi^-$ or may be derived using the Omnes representation
from the $\pi\pi$ elastic phase shifts in the vector channel --- here the representation for $F_V(m_{\pi\pi})$
derived in Ref.~\cite{Stollenwerk:IP,Guo:2008nc} is applied. The uncertainty in this form factor is negligible
compared to the experimental uncertainties, which are presented in detail in the next paragraph. Furthermore,
following Refs.~\cite{Venugopal:1998fq,Holstein:2001,Stollenwerk:IP}, we parameterize the term $P(s_{\pi\pi})$
as a real polynomial of first order: 
\begin{equation}
P(s_{\pi\pi}) = 1+\alpha s_{\pi\pi} \ .
\end{equation}
The parameter $\alpha$ can then be determined from a fit to the data, which is shown with the solid curve in
the lower panel of Fig.~\ref{fig:final}. The result for $\alpha = 0$ is shown as the dotted curve.

Tests for systematic uncertainties of $\alpha$ have been performed by varying in the analysis chain one by one
the conditions for the suppression of splitoffs, the cut on the missing mass of the $^3$He$\pi^+\pi^-$ system,
the condition on the probability of the kinematic fit, the method of subtracting the background from direct
multi-pion production, the $\eta\to\pi^+\pi^-\pi^0$ contribution and the model used for the acceptance
correction. Additionally, subsets of the data, collected with different experimental settings, allowed to
cross check the influences of luminosity variations and different RF settings during the measurement.

Using different methods to subtract the continuous multi-pion background, different models to perform the
acceptance correction or using different RF settings does not cause statistically significant deviations
from the original result. The contributions to the systematic uncertainty of $\alpha$ derived from the other
tests are listed in Tab.~\ref{tab:alpha_sys}. The overall systematic error is obtained by adding the
contributions quadratically.

One of the largest uncertainties results from the fluctuations of the luminosity during data taking. The
variation of the final result with the chosen luminosity can be explained by the accuracy of the simulations
concerning pile-up effects and beam-target overlap parameters, which have not been included in a systematic
way.

\begin{table}[hbt]
 \begin{center}
   \begin{tabular}{ c|c }
\hline
                          Test  &  $\sigma$ \\
\hline
                     Splitoffs  &  0.34 \\
        MM(He$\pi^+\pi^-$) cut  &  0.22 \\
          P($\chi^2_{kf}$,ndf)  &  0.12 \\
 $\eta\to\pi^+\pi^-\pi^0$  bkg. &  0.26 \\ 
           Luminosity  effects  &  0.32 \\
\hline
   \end{tabular}
 \end{center}
 \caption{\label{tab:alpha_sys}Summary of contributions to the systematic uncertainty of the parameter
$\alpha$. }
\end{table}%

Taking into account the systematic studies, the final result for the parameter $\alpha$ is:
$$\alpha = 1.89\pm0.25_{stat}\pm0.59_{sys}\pm0.02_{theo}\textrm{~GeV}^{-2}\qquad,$$
where the theoretical uncertainty results from the uncertainty of the pion vector form factor due
to the input of Ref.~\cite{GarciaMartin:2011} for the $\pi\pi$ phase shifts and  the extrapolation beyond the
upper cutoff. For more details, see Ref.~\cite{Stollenwerk:IP}.

In comparison to theory, calculations based on vector meson dominance~\cite{Picciotto:1991,Benayoun:2003}
result in a shape of the differential distribution corresponding to an \mbox{$\alpha =
(0.23\pm0.01)\,\textrm{GeV}^{-2}$}. The shape given by a parameterization of the pion vector form factor
combined with a fit to vector meson dominance~\cite{Venugopal:1998fq,Holstein:2001} corresponds to an
\mbox{$\alpha = (0.64\pm0.02)\,\textrm{GeV}^{-2}$}. The E$_\gamma$ spectrum from one--loop Chiral Perturbation
Theory~\cite{Bijnens:1989} can be described with an \mbox{$\alpha = -(0.7\pm0.1)\,\textrm{GeV}^{-2}$}. Thus,
the available theory descriptions produce distributions of E$_\gamma$, which are close to the curve of $\alpha
= 0$ shown with a dotted line in the right panel of Fig.~\ref{fig:final}. Within the total error of the
measurement, the value of $\alpha$ found in this work appears to be only compatible with the works of
Refs.~\cite{Venugopal:1998fq,Holstein:2001}.

\begin{table}[hbt]\small
  \begin{center}
    \begin{tabular}{ccc}
 $E_\gamma$ [GeV] & Entries [a.u.] & stat. [a.u.] \\
\hline
0.0225 &   97 & 653 \\
0.0275 &  407 & 394 \\
0.0325 &  530 & 302 \\
0.0375 & 1537 & 252 \\
0.0425 & 1291 & 220 \\
0.0475 & 2331 & 227 \\
0.0525 & 2397 & 218 \\
0.0575 & 2841 & 217 \\
0.0625 & 3604 & 228 \\
0.0675 & 4171 & 235 \\
0.0725 & 4887 & 247 \\
0.0775 & 5057 & 248 \\
0.0825 & 5091 & 252 \\
0.0875 & 6444 & 273 \\
0.0925 & 6673 & 282 \\
0.0975 & 6595 & 282 \\
0.1025 & 6448 & 284 \\
0.1075 & 6961 & 291 \\
0.1125 & 7051 & 297 \\
0.1175 & 7242 & 304 \\
0.1225 & 7138 & 300 \\
0.1275 & 7514 & 312 \\
0.1325 & 6963 & 309 \\
0.1375 & 7425 & 317 \\
0.1425 & 7014 & 319 \\
0.1475 & 6892 & 323 \\
0.1525 & 6600 & 324 \\
0.1575 & 5833 & 315 \\
0.1625 & 5739 & 320 \\
0.1675 & 4200 & 294 \\
0.1725 & 3942 & 292 \\
0.1775 & 2985 & 270 \\
0.1825 & 2334 & 244 \\
0.1875 & 1760 & 217 \\
0.1925 & 1250 & 207 \\
0.1975 & 384  & 125 \\
0.2025 & 63   &  79 \\
\hline
    \end{tabular}
   \end{center}
\caption{\label{tab:EGfinal}Distribution of the photon energy in the $\eta$ rest frame with statistical
errors. The values of E$_\gamma$ are central values of bins with a width of 5~MeV.}
\end{table}

\section{Conclusions}
For the first time, background subtracted and acceptance corrected differential distributions of the decay
$\eta\to\pi^+\pi^-\gamma$ have been extracted in the analysis of exclusive data. The distributions clearly
show the importance of final state interactions. The shape of the $E_\gamma$ spectrum can be very well
described by a parameterization that includes the factors required by gauge invariance and the centrifugal
barrier as well as the pion vector form factor times a first-order polynomial written as $(1+\alpha
s_{\pi\pi})$. A fit to the data gives \mbox{$\alpha =
1.89\pm0.25_{stat}\pm0.59_{sys}\pm0.02_{theo}\textrm{~GeV}^{-2}$}. In
order to shed further light on the anomalous sector of QCD, future theoretical studies will have to explain
simultaneously both the value of $\alpha$ as well as the branching ratio for $\eta\to\pi^+\pi^-\gamma$.

In recent production runs of the WASA facility at COSY further data on $\eta$ decays have been taken with high
statistics. From a preliminary analysis at least an order of magnitude more fully reconstructed
$\eta\to\pi^+\pi^-\gamma$ events is expected. The analysis of the acquired data will significantly decrease
not only the statistical but also the systematic uncertainties by an improved understanding of background
contributions.

The data will also be used to determine the branching ratio of the decay $\eta\to\pi^+\pi^-\gamma$. A recent
measurement of the CLEO collaboration~\cite{Lopez:2007} shows a relative branching ratio which differs by
more than three standard deviations from the results of previous measurements~\cite{Gormley:1970,
Thaler:1973}. Due to the unbiased tagging of $\eta$ mesons in the reaction \mbox{pd$\to ^3$He~$\eta$} it is
not only possible to extract relative but also absolute branching ratios at the WASA facility. This will help
to resolve the discrepancy.

\section{Acknowledgments}

We thank F.-K. Guo for providing us with the code to calculate the pion vector form factor.

This  work was in part supported  by: the  Forschungszentrum J\"ulich including the  JCHP-FFE program, the
European Commission under the 7th Framework Programme through the 'Research Infrastructures' action of the
'Capacities' Programme. Call: FP7-INFRASTRUCTURES-2008-1, Grant Agreement N. 227431,  the German  BMBF, the
German-Indian  DAAD-DST exchange  program, VIQCD  (VH-VI-231), the German Research Foundation (DFG), and the
Polish National Science Centre and Foundation for Polish Science - MPD program, co-financed by the European
Union within the European Regional Development Fund.

We gratefully acknowledge the financial support given by the Knut and Alice Wallenberg Foundation, the Swedish
Research Council, the G\"oran Gustafsson Foundation, the Polish Ministry of Science and Higher Education   
under the grant PBS 7P-P6-2/07.

We also want to thank the technical and administrative staff at the Forschungszentrum J\"ulich, especially at
the COoler SYnchrotron COSY and at the participating institutes.

This work is part of the PhD Thesis of C. F. Redmer.

\end{document}